\documentclass{article}
\usepackage[utf8]{inputenc}
\usepackage{graphicx,wrapfig,lipsum} %use Italian accents
\usepackage[T1]{fontenc}    % use 8-bit T1 fonts
\usepackage{hyperref}       % hyperlinks
\usepackage{url}            % simple URL typesetting
\usepackage{booktabs}       % professional-quality tables
\usepackage{amsfonts}       % blackboard math symbols
\usepackage{nicefrac}       % compact symbols for 1/2, etc.
\usepackage{microtype}      % microtypography
\usepackage{siunitx}
\DeclareSIUnit{\sample}{Sa}
\usepackage{color}
\usepackage{float}  
\usepackage{amsmath}
\definecolor{dkgreen}{rgb}{0,0.6,0}
\usepackage{setspace}
 
\definecolor{gray}{rgb}{0.5,0.5,0.5}
\definecolor{mauve}{rgb}{0.58,0,0.82}
\usepackage{subcaption}
\usepackage[utf8]{inputenc} % allow utf-8 input  %use Italian accents
\usepackage[T1]{fontenc}    % use 8-bit T1 fonts
\usepackage{hyperref}       % hyperlinks
\usepackage{url}            % simple URL typesetting
\usepackage{booktabs}       % professional-quality tables
\usepackage{amsfonts}       % blackboard math symbols
\usepackage{nicefrac}       % compact symbols for 1/2, etc.
\usepackage{microtype}      % microtypography
\usepackage{siunitx}
\DeclareSIUnit{\sample}{Sa}
 
\usepackage[table]{xcolor}
\usepackage{geometry}
\usepackage{tikz}
\usetikzlibrary{cd}
\usetikzlibrary{tikzmark}
\usepackage{physics}
\usepackage{listings}
\usepackage{caption}
\usepackage{subcaption}
\usepackage{amssymb}
\usepackage{enumerate}
\usepackage{comment}
\usepackage{mathrsfs}
\usepackage{wrapfig}
\usepackage{chngcntr}
\usepackage{courier}
\usepackage[euler]{textgreek}

\usepackage{wrapfig}
\usepackage{slashed}
\usepackage{array}

\usepackage{tikz}
\usetikzlibrary{cd}
\usetikzlibrary{tikzmark}
\usepackage{physics}
\usepackage{listings}
\usepackage{amssymb}
\usepackage{enumerate}
\usepackage{comment}
\usepackage{mathrsfs}

\usepackage{chngcntr}
\usepackage{courier}
\usepackage[euler]{textgreek}
\usepackage{verbatim}
\usepackage{blindtext}
\usepackage{amsmath}
\usepackage{imakeidx}
\makeindex[columns=3, title=Alphabetical Index, intoc]
\usepackage{setspace}
\usepackage{array}

\makeatletter
\newcommand{\thickhline}{%
    \noalign {\ifnum 0=`}\fi \hrule height 1pt
    \futurelet \reserved@a \@xhline
}
\newcolumntype{"}{@{\hskip\tabcolsep\vrule width 1pt\hskip\tabcolsep}}
\makeatother

\counterwithin{figure}{section}
\counterwithin{table}{section}
\counterwithin{equation}{section}

\begin{document}
    
\thispagestyle{plain}
\begin{center}
    {\Huge DARK MATTER IMPLICATIONS}
    \vspace{0.2cm}
    {\Huge OF THE NEUTRON ANOMALY}
        
    \vspace{1.6cm}
    
    \vspace{0.4cm}
    Leonardo Darini
       
    \vspace{0.3cm}
    \emph{Dipartimento di Fisica, Università di Pisa, Italia}

    \vspace{0.2cm}
\end{center}
    Motivated by the neutron decay anomaly, we reconsider the neutron decay model $n \rightarrow \chi\chi\chi$, where the new species $\chi$ plays the role of dark matter. 
    We precisely compute the $\bar{\chi} n \rightarrow \chi\chi$ rate finding that fitting the anomaly compatibly with all bounds needs two ''generations'' of the $\chi$ particle.
\tableofcontents
%%%%%%%%%%%%%%%%%%%%%%%%%%%%%%%%%%%%%%%%%%%%%%%%%%%%%%%%%%%%
\section{Introduction}
The current measuraments of the neutron life-time present an inconsistency. Two methods are employed and give different values. One called \emph{beam method} measures the $\beta^-$ decay rate, $\Gamma^{\beta}_{n} = 1/(888 \pm 2 \,\, \text{s})$, by counting the protons produced from a beam of cold neutrons \cite{beam1,beam2}. The other called \emph{bottle method} measures the total neutron decay width, $\Gamma^{tot}_{n} = 1/(878.3 \pm 0.3 \,\, \text{s})$, by storing ultra-cold neutrons in a magnetic bottle \cite{bottle1,bottle2,bottle3}.
These measuraments are in disagreement at $\sim 4.6 \sigma$ with the Standard Model, that predicts $\Gamma^{tot}_n = \Gamma^{\beta}$. It has been proposed that $\Delta \Gamma = \Gamma^{tot}_n - \Gamma^{\beta} \approx 1.2 \times 10^{-5} (1 \pm 0.21) \frac{1}{\text{sec}}$ can be due to an extra decay channel of the neutron. The new decay needs to be nearly invisible with a branching ratio around 1\%.
Various authors proposed a $n \rightarrow \chi \gamma$ \cite{FG1} decay into a new neutral fermion $\chi$ with mass $M$ slightly below the neutron mass so that $E_{\gamma} \approx m_n - M$ is small. This new particle $\chi$ does not decay back to SM charged particles if its mass is $M < m_p + m_e$ and can thus be considered as a dark matter candidate, a possibility disfavoured by tests \cite{Failure1FG,Failure2FG} which don't see the final photon with the predicted energy. Furthermore, the $\chi$'s thermalize inside neutron stars (NS) \cite{Failure3FG,Failure4FG} softening the equation of state\footnote{This problem can be avoided by adding new light mediator such that $\chi$ undergoes repulsive interactions stronger than the QCD repulsion among neutron.} too much, reducing the NS mass ($M_{NS}$) below $0.7 M_{\odot}$ while $ M_{NS} \sim 2M_{\odot}$ is observed.
In this perspective we consider a different model that was proposed in \cite{Strumia}. The new decay is $n \rightarrow \chi\chi\chi$. In the more minimal model $\chi$ carries baryon number $B_{\chi} = 1/3$. The chemical potential of $\chi$ in thermal equilibrium in a neutron star is fixed in terms of the chemical potential of the conserved charges as $\mu_{\chi} = B_{\chi} \mu_{n}$. This ensures a substantial reduction of the impact on neutron stars. In section 2, we discuss the main features of the dark species. In section 3, we show that all the bounds are satisfied if more than one $\chi$ species are considered. In section 4, we investigate direct detections of DM.
In section 5, we show the equations of state of the neutron star resulting from calculations for different number of generation $N$ of the particle $\chi$ compared with predictions from the SM. 
The conclusions are presented in the last section.
%%%%%%%%%%%%%%%%%%%%%%%%%%%%%%%%%%%%%%%%%%%%%%%%%%%%%%%%%%%%
\section{Model}
The particle $\chi$ must be light enough to allow the decay $n \rightarrow \chi\chi\chi$, so $M \lesssim m_n/3$. 
The DM mass $M$ is strongly constrained by the following thresholds. Proton stability gives $M > (m_p - m_e)/3 = 312.59$ MeV. Hydrogen decay $H \rightarrow \chi\chi\chi\nu_e$ is kinematically open for $M < (m_p + m_e)/3 \approx 312.93$ MeV. The bound from Beryllium ($^8\text{Be}$) nuclear decay implies $M < (m_n - E_{\text{Be}})/3 \approx 312.63$ MeV where $E_{\text{Be}} = 1.664$ MeV \cite{FG1}.
The number of generations $N$ for this new particle will be important when making comparisons with the experimental bounds. If one assumes more generations of $\chi_{i}$ the NS mass remains close enough to the SM limit \cite{Strumia}. In particular, considering the behaviour of NS mass 
as function of NS radius, $M_{NS}$ undergoes only a small softening, thus remaining compatible with the data. Different SM computations lead to maximal NS masses between $1.8M_{\odot}$ and $2.6M_{\odot}$ and minimal radii between $10\, \text{km}$ and $14 \,\text{km}$, in apparent agreement with the data.
We compute the $N=1$ and $N=3$ cases.
The theory employed is the following, with $n$ the neutron field and $\Psi$ the dark particle field:
\begin{equation}\label{lagrangianN1}
   \mathscr{L}_{\text {eff }}=\mathscr{L}_{\mathrm{SM}}+\bar{\Psi}(i \slashed{\partial}-M) \Psi+\frac{\left(\bar{\Psi}^{c} \Gamma \Psi\right)(\bar{n} \Gamma \Psi)+\text { h.c. }}{3 ! \Lambda_{\chi n}^{2}}.
\end{equation}
The new interaction term is \emph{unusual} because we want three final particles and no anti-particles.
Following \cite{Strumia} we consider the vectorial-axial and scalar interactions. We study in detail the cases with $N=3$ generations which will prove to be in agreement with the data. 
%%%%%%%%%%%%%%%%%%%%%%%%%%%%%%%%%%%%%%%%%%%%%%%%%%%%%%%%%%%%
\section{The Neutron Decay Anomaly}
In this section we study the lifetime of the free neutron in the case of $N=3$ generations of the $\chi$ particle. 
This will be important in the context of direct detection later.
In the previous papers, e.g. \cite{Strumia}, were given only estimatations of $N>1$ cases. First we calculate
the decay widths of the free neutron and subsequently we calculate 
the lifetimes for various processes involving the neutron. 
The $n \rightarrow \chi\chi\chi$ decay with $N=1$ arises from 4-fermion effective operators in Eq. (\ref{lagrangianN1}). The decay widths for scalar left-right and vector-axial couplings were calculated in \cite{Strumia}:
\begin{equation}\label{N1widths}
   \Gamma_{n \rightarrow \chi\chi\chi} = \frac{m_{n}^{5}}{27 \pi^{3} \Lambda_{\chi n}^{4}}\left\{\begin{array}{ll}g_{L}^{2} g_{R}^{2}\left(1-3 M / m_{n}\right)^{3} / 16 & \text { if } \Gamma^{L-R}=g_{L} P_{L}+g_{R} P_{R} \\ g_{A}^{2}\left(1-3 M / m_{n}\right)^{3} & \text { if } \Gamma^{V-A}=\gamma_{\mu}\left(g_{V}+g_{A} \gamma_{5}\right)\end{array}\right.
\end{equation}
The Lagrangian\footnote{One choice of the i,j,k indices per dark particle is arbitrarily fixed, the combinatoric factor 1/3! is excluded.} for the $N=3$ generation process is:
\begin{equation}\label{lagrangianN3}
   \mathscr{L}_{\text {eff }}=\mathscr{L}_{\mathrm{SM}}+\sum_{f=1}^{3}\bar{\Psi}_f(i \slashed{\partial}-M) \Psi_{f}+\frac{\left(\bar{\Psi}^{c}_3 \Gamma \Psi_2 \right)(\bar{n} \Gamma \Psi_1)+\text { h.c. }}{ \Lambda_{\chi n}^{2}}.
\end{equation}
The free neutron decay widths using the scalar and vectorial\footnote{If we consider the pure axial coupling we obtain the same expression by replacing $g_V$ with $g_A$.} couplings are:
\begin{equation}\label{widthN3leftright}
   \Gamma^{L-R, N=3}_{n \rightarrow \chi \chi \chi} = \frac{m_n^5 \left(g_L^2-g_R^2\right)^2 \left(1-\frac{
     3 M}{m_n}\right)^2}{27 \times 64 \pi ^3 \Lambda
      _{\text{$\chi $n}}^4}, \qquad \Gamma_{n \rightarrow \chi \chi \chi}^{V, N=3}= \frac{5 g_V^4 m_n^4 \left(1-\frac{3 M}{m_n}\right)^2}{27 \times 24 \pi
   ^3 \Lambda _{\chi n}^4}. 
\end{equation}
The decay width in the $N=1$ case is proportional to $\epsilon^3$, with $\epsilon = \left(1-\frac{3 M}{m_n}\right)$. In the case $N>1$ the width is $\propto \epsilon^2$. The bounds on $ \Gamma_{n \rightarrow \chi \chi \chi \gamma}$, $   \Gamma_{\mathrm{H} \rightarrow \chi \chi \chi \nu_{e}}$ and $\Gamma_{\mathrm{H} \rightarrow \chi \chi \chi \nu_{e} \gamma}$ are satisfied in both cases \cite{Strumia}. 
%%%%%%%%%%%%%%%%%%%%%%%%%%%%%%%%%%%%%%%%%%%%%%%%%%%%%%%%%%%%
\vspace{+2em}
\section{Direct Detection of Dark Matter and Cosmology}
The neutron anomaly is reproduced for $\Lambda_{\chi n} \gg v$ so the interaction rates of $\chi$ particles at $T \sim m_n$ are below electroweak rates. The $\chi$ species can be produced by freeze-in \cite{Strumia}.
The $\bar{\chi} n \leftrightarrow \chi\chi$ process leads to unusual DM direct detection signals. It is kinematically open today and the non-relativistic $\bar{\chi}$ and $n$ in its initial state produce relativistic $\chi$ with energy $E = 2m_n/3$ \cite{Strumia}.
In this section we compute its rate and compare with bounds.
%%%%%%%%%%%%%%%%%%%%%%%%%%%%%%%%%%%%%%%%%%%%%%%%%%%%%%%%%%%%
\subsection{Signals from $\bar{\chi}n \rightarrow \chi\chi$ process}
\begin{figure}[H]
   \centering
   \includegraphics[width=11cm]{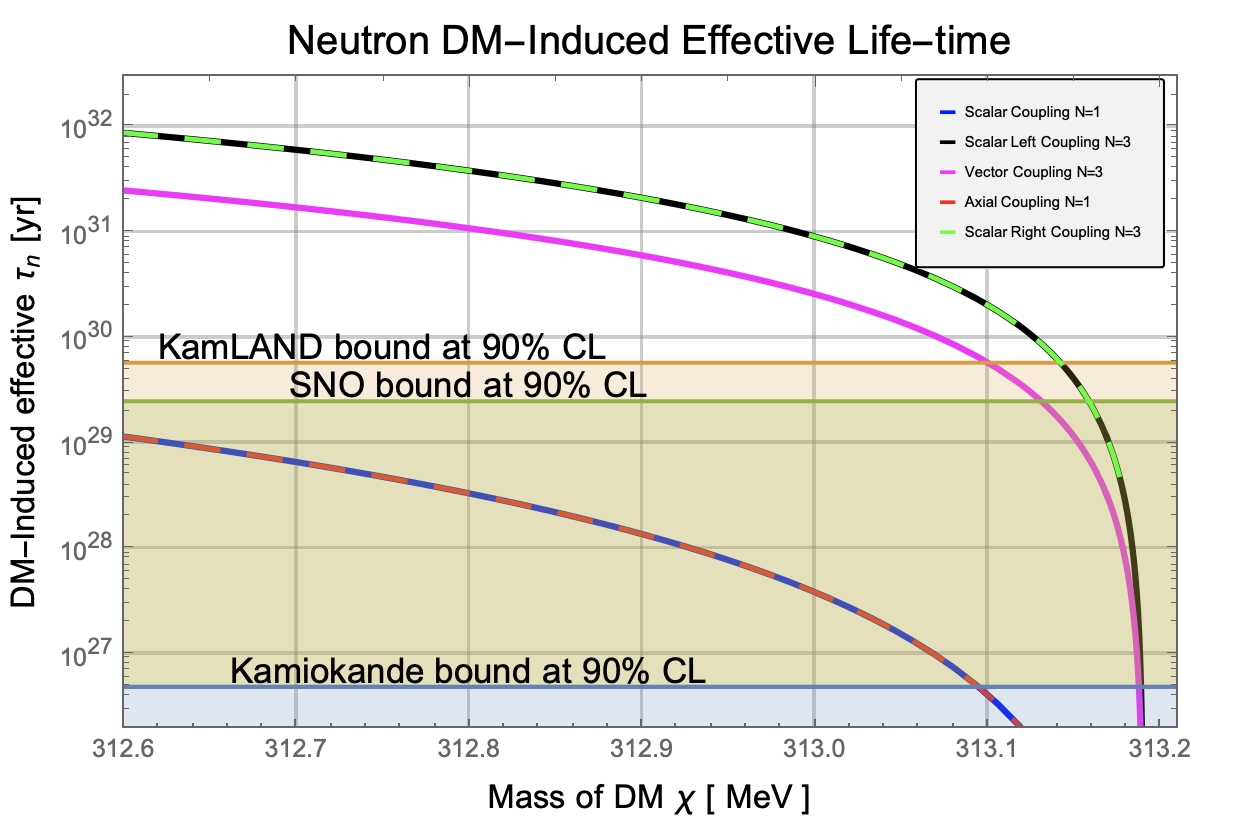}
   \caption{\emph{Plot of effective $\tau_n$ as a function of the mass of the dark particle $\chi$ in the allowed kinematic range. We show the two different cases: for N=1, excluded (SNO and KamLand bounds) and for N=3 above the bounds}}
   \label{fig:3}
\end{figure}

We consider the scalar (left-right) coupling, $\Gamma = g_LP_L +g_R P_R$. 
Later, calculations with the vector-axial coupling will also be considered.
First, we set $N=1$ generations for $\chi$ and give a calculation of cross section.
In particular, for this process, the initial states are non-relativistic and the products of scattering are relativistic.
The prediction for $\sigma v_{rel}$ is:
\begin{equation}\label{crossectionN1}
    \sigma^{N=1,L-R}_{\bar{\chi} n \rightarrow \chi \chi} v_{rel} =  \frac{\text{g}_L^2 \text{g}_R^2 m_n^2}{96 \sqrt{3}\pi \Lambda_{ \chi n}^4}.
\end{equation}
Fixing $\Delta \Gamma$ from the free neutron decay rate gives the cross section as a function of the $\chi$ mass.
The cross section is well below current bounds from direct detection experiments for the value of $\Lambda_{\chi n}$ motivated by the neutron decay anomaly.
However different experiments are more sensitive.
\begin{comment}
\begin{figure}[h]
    \centering
    \includegraphics[width=12cm]{crosssection.png}
    \caption{\emph{Predicted $\sigma v_{rel}$ for the production of two $\chi$ with scalar coupling and $N=1$ on a semilogarithmic scale.}.}
    \label{fig:2}
 \end{figure}
\end{comment}
%%%%%%%%%%%%%%%%%%%%%%%%%%%%%%%%%%%%%%%%%%%%%%%%%%%%%%%%%%%%
Dark matter scattering makes ordinary matter radioactive since a neutron that disappears within a nucleus leaves a hole, triggering nuclear deexcitations and decays\footnote{DM scatterings which convert $^{16}\text{O}_{8}$ into $^{15}\text{O}^{*}_{8}$ emitting a photon are a class of nuclear decay induced by dark matter.}.
The expression for effective lifetime is
\begin{align}
   &\tau_{n}^{\text{eff}}= \left(\frac{\rho_{\odot}}{2 M} \sigma_{\bar{\chi} n \rightarrow \chi \chi}v_{rel} \right)^{-1}.
\end{align}
Bounds on the neutron effective lifetime are found in various experiments:
\begin{equation}\label{bounds}
   \tau(n \rightarrow \text{invisible})>\left\{\begin{array}{lll}4.9 & 10^{26} \mathrm{yr} & \text { from KAMIOKANDE } \cite{Kamiok} \\ 2.5 & 10^{29} \mathrm{yr} & \text { from SNO } \cite{SNO} \\  5.8  &  10^{29} \mathrm{yr} & \text { from KAMLAND }\cite{KamLand} \end{array}\right.
\end{equation}
The DM-induced neutron lifetime satisfies the bounds in Eq. (\ref{bounds}), if the number of $\chi$ generations are larger than one. So, in this perspective we fix $N=3$.
We take into account the lagrangian in Eq. (\ref{lagrangianN3}).
The expressions for the cross sections per relative velocity are:
\begin{equation}
   \sigma^{L,N=3}_{\bar{\chi} n \rightarrow \chi \chi}v_{rel}  = \frac{7 g_L^4 m_n^2}{24 \times 16 \sqrt{3} \pi \Lambda_{ \chi n}^4}, \qquad \sigma^{V,N=3}_{\bar{\chi} n \rightarrow \chi \chi}v_{rel}  = \frac{40 g_V^4 m^2_n}{3 \times 16 \sqrt{3} \pi  \Lambda _{\chi n}^4}.
\end{equation}
In previous expressions the density of DM is $\rho_{\odot} = 0.4 \, \text{GeV}/ \text{cm}^3$. The $N=1$ case calculations gives equal expressions for the vectorial coupling and the scalar coupling. $\Gamma(n \rightarrow \chi\chi\chi)$ is fixed from the anomaly.
We show our results in Fig. \ref{fig:3}.
The plot shows that the $N=3$ case satisfies all the bounds. The predictions for $N=1$ are excluded by SNO and by KamLand.
The main reason for this difference is the proportionality to different powers of the $\epsilon = \left(1-\frac{3 M}{m_n}\right)$ factor. $\Gamma(n \rightarrow \chi\chi\chi)$ in the $N=1$ case is $ \propto \epsilon^3$ whereas in the $N=3$ case it is $ \propto \epsilon^2$.
Since $\Gamma(n \rightarrow \chi\chi\chi)$ for $N>1$ are $\propto \epsilon^2$ all these cases satisfy bounds.
%%%%%%%%%%%%%%%%%%%%%%%%%%%%%%%%%%%%%%%%%%%%%%%%%%%%%%%%%%%%
\section{Neutron Star Physics}
Neutron stars are described by the Tolmann-Oppenheimer-Volkoff (TOV) equations. 
\begin{equation}\label{TOVeq}
   \frac{d \wp}{d r}=-\frac{G}{r^{2}} \frac{\left(\mathscr{M}(r)+4 \pi r^{3} \wp\right)(\rho+\wp)}{1-2 G \mathscr{M}(r) / r}, \qquad \frac{d \mathscr{M}}{d r}=4 \pi r^{2} \rho.
\end{equation}
These equations can be solved for any equation of state that gives the density $\rho$ in terms of the pressure $\wp$, starting with an arbitrary pressure at the center, $r = 0$, where $\mathscr{M}(0) = 0$, and by evolving outwards up to the neutron star radius $r = R$ at which $\rho(R) = 0$. 
In the same way one can predict the relation between the radius $R$ and the total mass $\mathscr{M}$.
We add new particles $\chi$ to neutron and to sub-dominant SM particles, in order to have $\rho = \rho_{n}+ \rho_{\chi}$ and $\wp = \wp_{n}+ \wp_{\chi}$.
The energy density and the pressure are
\begin{equation}\label{integrali}
  \rho_{\chi}=g \int \frac{d^{3} p}{(2 \pi)^{3}} E, \qquad \wp_{\chi}=g \int \frac{d^{3} p}{(2 \pi)^{3}} \frac{p^{2}}{3 E}.
\end{equation}
Thermal equilibrium of $n \leftrightarrow \chi\chi\chi$ relates the chemical potential of $\chi$, $\mu_{\chi} = \sqrt{M^2 +p_{\chi}^2}$ to the
chemical potential of neutrons, $\mu_{\chi} = \mu_n/3$.
In our calculations we adopt the \emph{BSk24} from \cite{BSkS24}.

\subsection{Equations of State}
\begin{figure}[H]
   \centering 
   \includegraphics[width=0.5\textwidth]{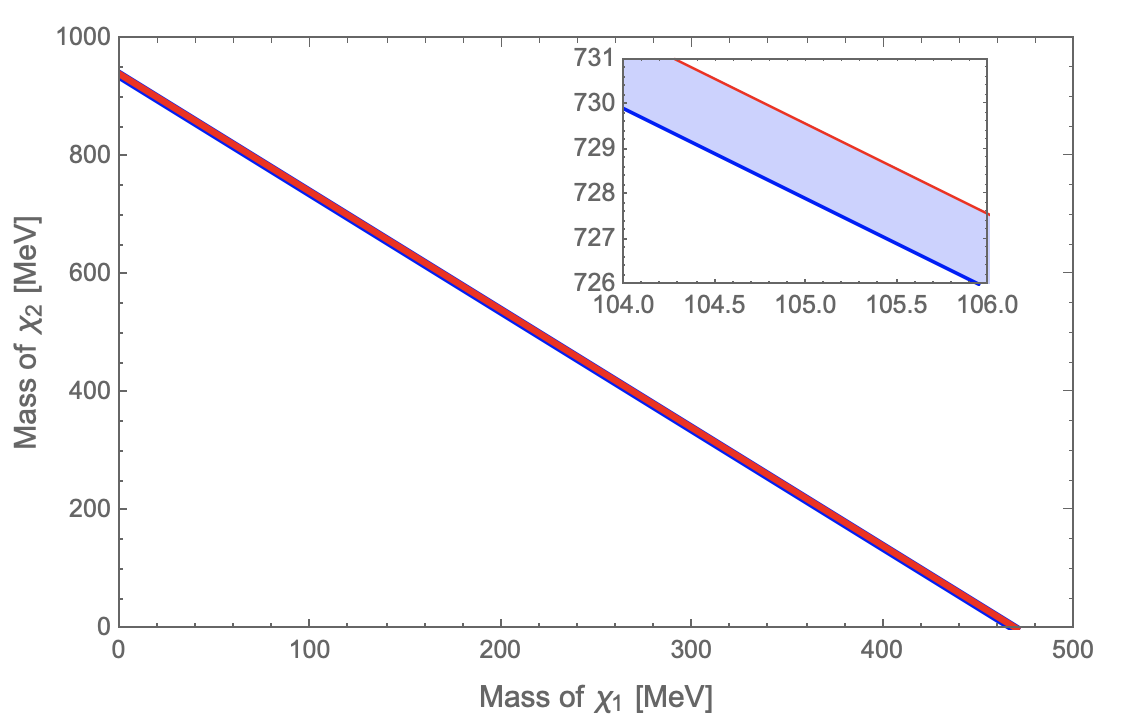}
   \caption{\emph{Allowed mass range. The subplot shows a magnification of an arbitrary range.}}
   \label{fig:range}
\end{figure}

\begin{figure}[H]
   \centering
   \begin{subfigure}[b]{0.45\textwidth}
       \centering
       \includegraphics[width=\textwidth]{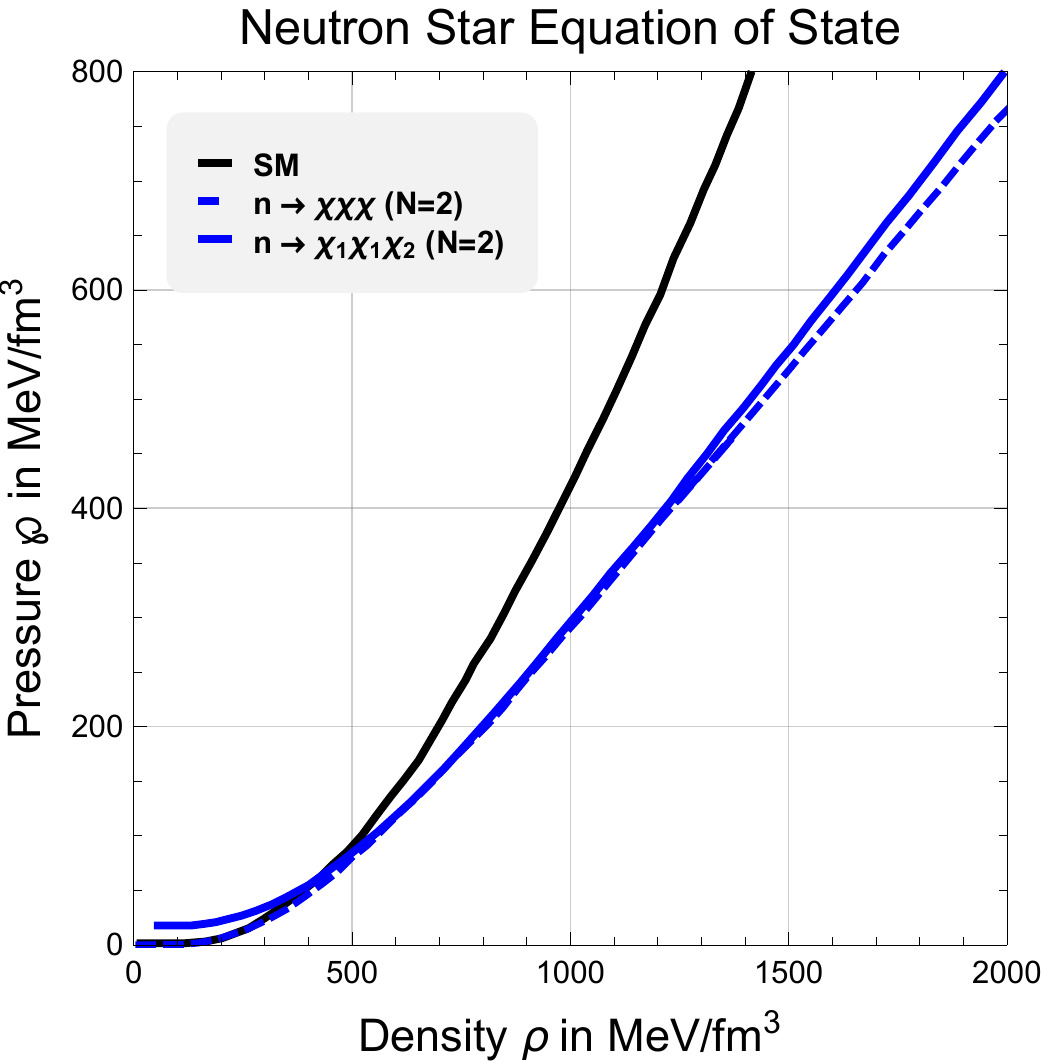}
       \label{fig:Eosmodif}
   \end{subfigure}
   \hfill
   \begin{subfigure}[b]{0.45\textwidth}
       \centering
       \includegraphics[width=\textwidth]{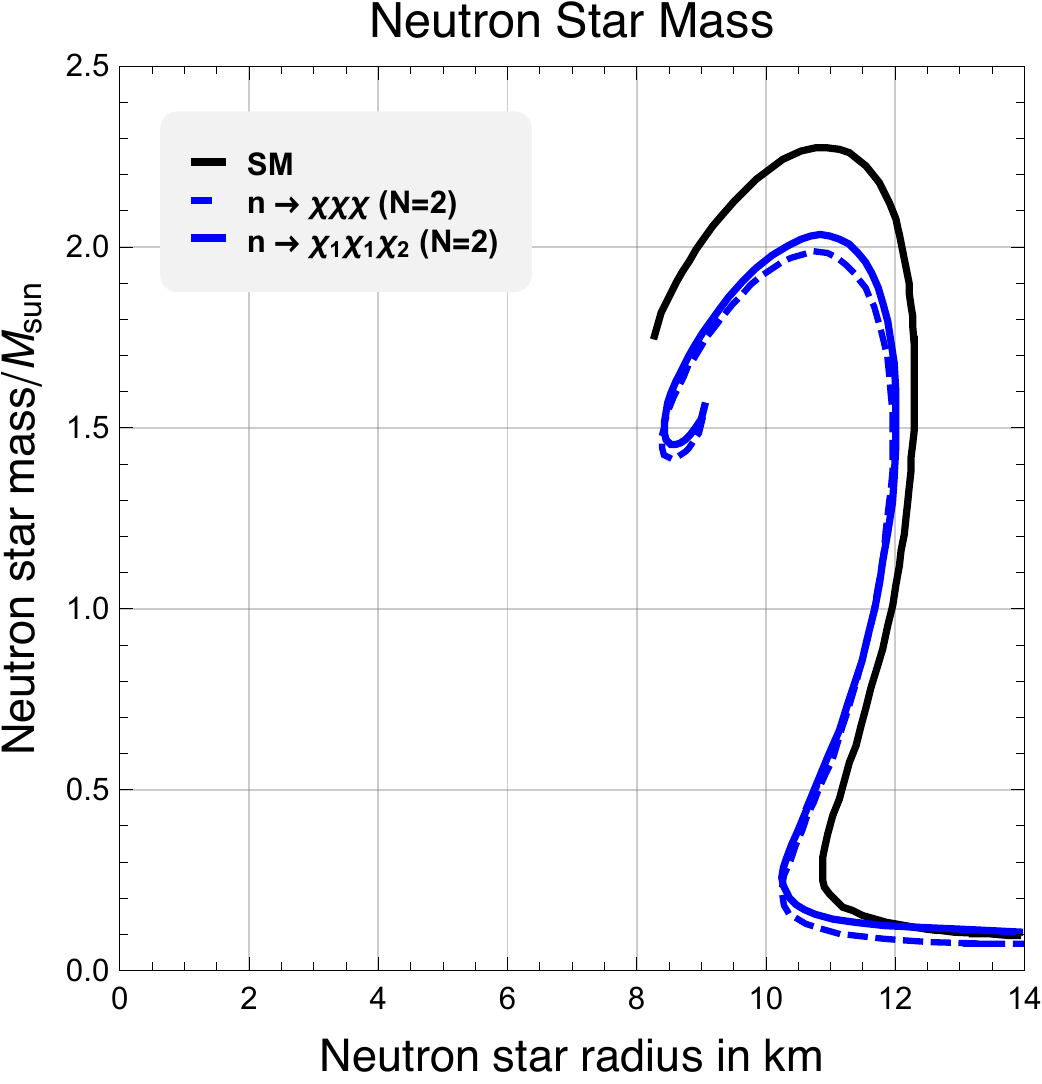}
       \label{fig:Massmodif}
   \end{subfigure}
   \caption{\emph{Equations of state for neutron stars considering the BSk24 equation of state from SM \cite{BSkS24}. \textbf{Left}.
   For the SM (black curve); $n \leftrightarrow \chi_1\chi_1\chi_2$ for N=2 case (blue curves); $n \leftrightarrow \chi\chi\chi$ for N=2 case in the limit of equal masses of $\chi$s (red curves). \textbf{Right}. The corresponding relationship
  between the radius and mass of neutron stars.}}
   \label{fig:stelledineutronimodif}
\end{figure}

\begin{figure}[H]
   \centering
   \begin{subfigure}[b]{0.45\textwidth}
       \centering
       \includegraphics[width=\textwidth]{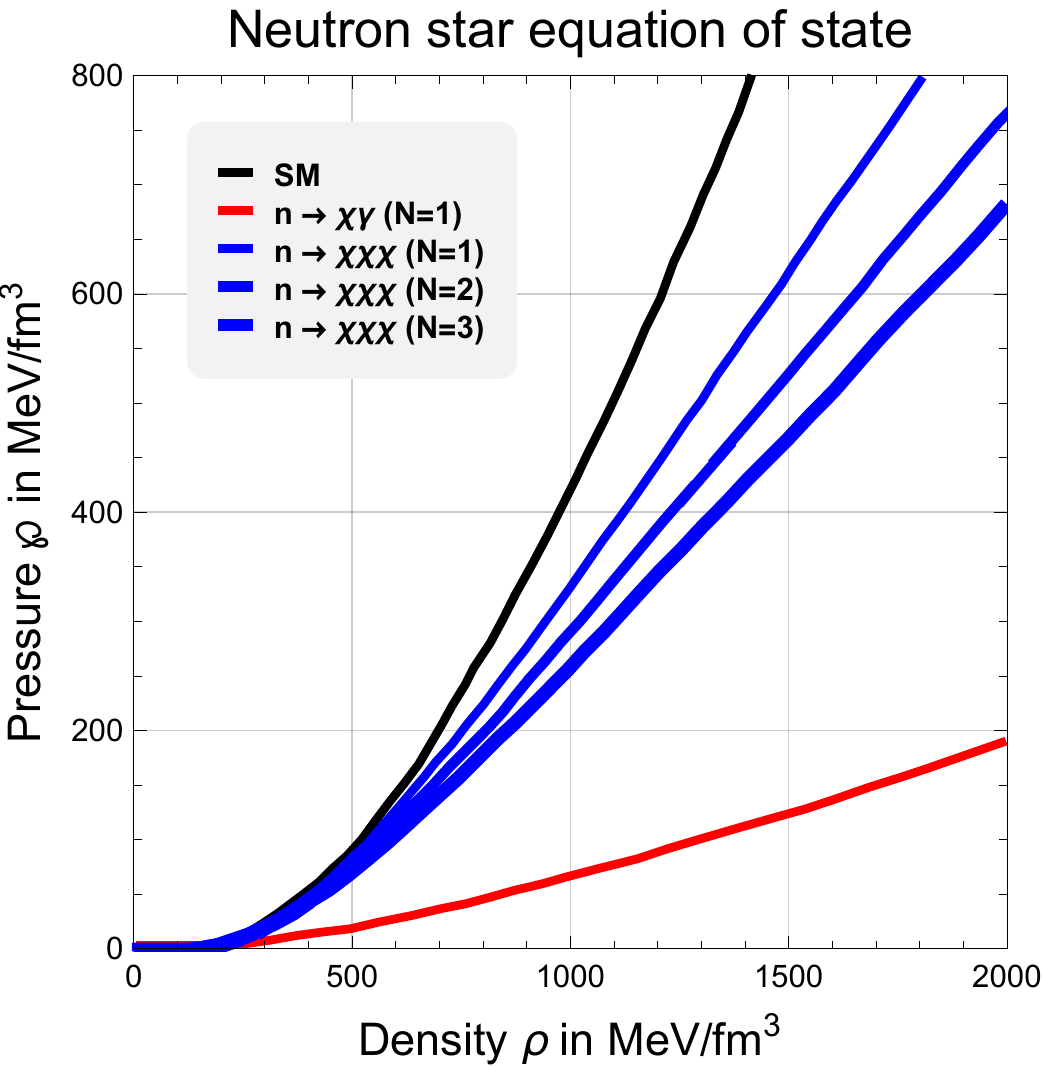}
       \label{fig:Eos}
   \end{subfigure}
   \hfill
   \begin{subfigure}[b]{0.45\textwidth}
       \centering
       \includegraphics[width=\textwidth]{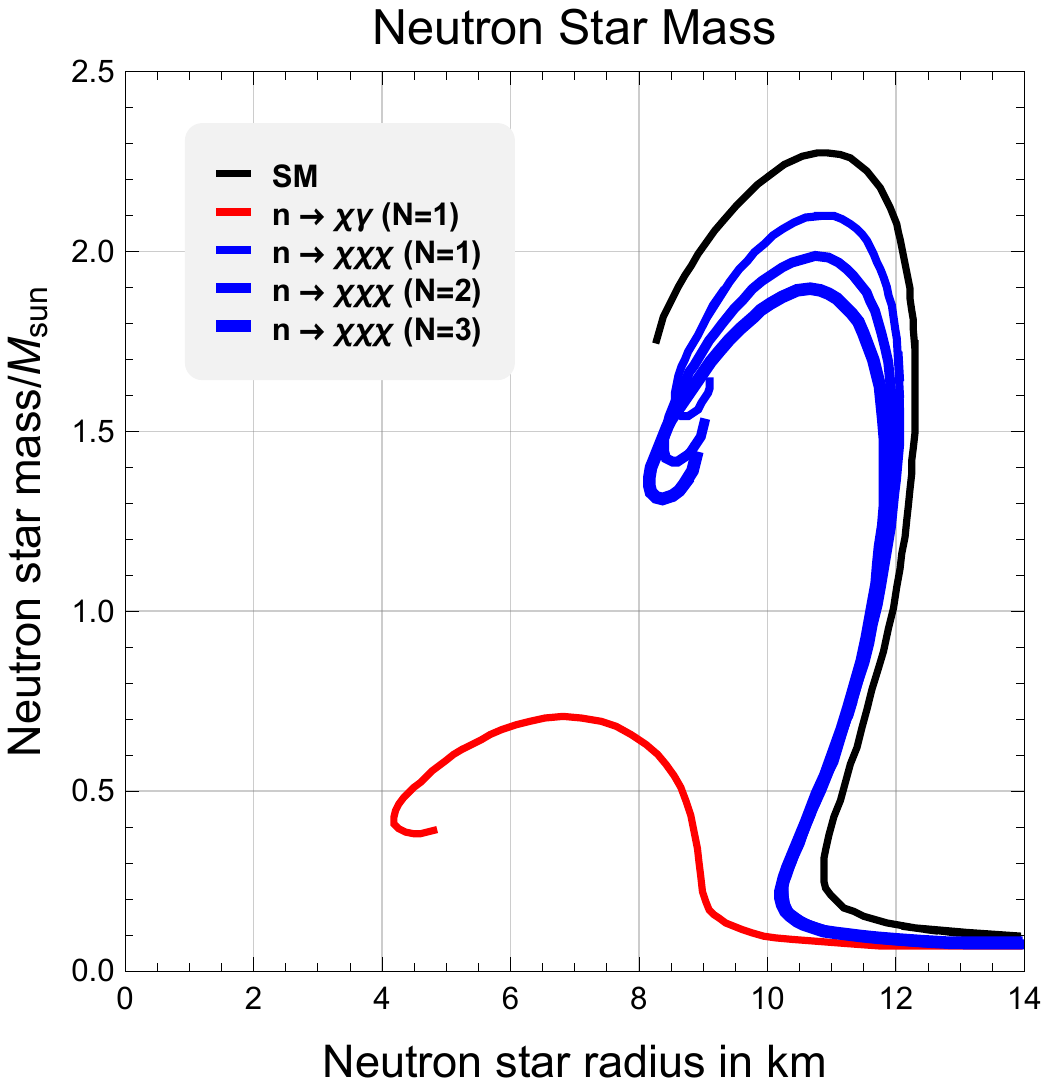}
       \label{fig:Mass}
   \end{subfigure}
   \caption{\emph{\textbf{Left.} Equations of state for neutron stars considering the BSk24 equation of state from \cite{BSkS24}. In the SM (black curve); $n \leftrightarrow \chi\chi\chi$ for different number of generations N (different blue curves); $n \leftrightarrow \chi\gamma$ in N=1 (red curves). 
   \textbf{Right.} The corresponding relationship between the radius and mass of neutron stars. This shows that the observed NS with masses around two solar masses are compatible with $n \leftrightarrow \chi\chi\chi$, but not with $n \leftrightarrow \chi\gamma$. The solutions below the peaks at smaller radii are unstable.}}
   \label{fig:stelledineutroni}
\end{figure}

We investigate the case in which the number of generations is $N = 2$ and the decay involves two different dark particles, $\chi_1$ and $\chi_2$. The decay is $n \leftrightarrow \chi_1\chi_1\chi_2$.
The kinematic threshold imposed by Beryllium is taken as the lower limit for the sum of the masses of the dark particles. Thus, we obtain $m_n > 2M_1 + M_2 > (m_n - E_{Be})$, where $E_{Be} = 1.664\,\, \text{MeV}$. The plot in Fig. \ref{fig:range} shows the allowed mass range.
We consider the baryonic number $B_{\chi_1} = B_{\chi_2} = 1/3$ and the chemical potential $\mu_{\chi_1} = \mu_{\chi_2} = \mu_n/3$. 
Given the two species at thermal equilibrium, we consider their contributions separately.
Now the energy density is $\rho = \rho_{SM} + \rho_{\chi_1} + \rho_{\chi_2}$.
Considering the allowed mass value: $M_1 \sim 1-100 \,\,\text{MeV}$ and $M_2 \sim 0.75-0.939 \,\, \text{GeV}$ one can obtain an EoS close to the prediction of the SM. 
The prediction of the SM is compared with that of the neutron decay for a number of generations $N = 2$ in the case with different masses of $\chi$ and in the limit of equal masses.
The results are shown in Fig. \ref{fig:stelledineutronimodif}. 
In Fig. \ref{fig:stelledineutroni} we show the results of the TOV equations for different models. 
The red curve in the plot on the left shows how the equations of state calculated in the SM soften too much if $n \leftrightarrow \chi \gamma$ is in thermal equilibrium (i.e. $\mu_n = \mu_{\chi}$) with $M = m_n$. 
The blue curves in the plot on the left show how thermal equilibrium of $n \leftrightarrow \chi\chi\chi$ (i.e. $\mu_{\chi}= \mu_n/3$), considering the limit of equal masses $M = m_n/3$ and for different number of generations $N = 1,2,3$, leads to a milder change in the equation of state. 
As the number of generations increases, the curve undergoes an increasing softening compared to the result in the SM. 
The curve scales approximately linearly with $N$.
Consequently, in the plot on the right we show that the relation between the neutron star mass $\mathscr{M}$ and radius $R$ in presence of $n \leftrightarrow \chi\chi\chi$ remains close to the SM limit (in particular taking into account the observed neutron stars with mass $\mathscr{M} \approx 2M_{\odot}$) even when the number of generations increases, in contrast with what happens if $n \leftrightarrow \chi\gamma$ is in thermal equilibrium. 
In particular, while $n \leftrightarrow \chi\gamma$ reduces the maximum mass of neutron stars in contradiction to the data, $n \leftrightarrow \chi\chi\chi$ leads to a slight reduction comparable to the current SM uncertainties. 
The radius of the neutron star is reduced equally slightly and is compatible with the data. Various calculations in the SM lead to maximum neutron star masses between $1.8 M_{\odot}$ and $2.6 M_{\odot}$ and minimum radii between $10$ and $14$ $km$ \cite{massradii}, in apparent agreement with the data. 
Therefore, if the SM is able to explain the observed neutron stars, so is its $n \rightarrow \chi\chi\chi$ extension. More precise future calculations and observations may be able to verify this slight difference.

\section{Conclusions}
In this work we investigated the decay model $n \rightarrow \chi_i\chi_j\chi_k$ for different number of generations of the species $\chi$ to solve the neutron lifetime anomaly. The results are in agreement with the experimental bounds from neutron physics.
In Section 3 we have shown the neutron decay rates for different couplings in the case $N=3$. This implies a fully invisible Hydrogen decay rate compatible with the inverse universe age and with the bound from the Borexino experiment.
In Section 4 we calculated the cross section and effective lifetimes for the process $\bar{\chi} n \rightarrow \chi\chi$, showing that the case $N=3$ satisfies the bounds from Kamiokande, SNO and KamLand.
In Section 5 we studied the equations of state for different number of generations of $\chi$ species.
We proposed a model for the $N=2$ case considering two different species $\chi_1$ and $\chi_2$ with different masses, and we showed that it produces an enhancement of the EoS respect to the prediction in the case with equal masses.
The equation of state $\wp(\rho)$ of neutron stars shown in Fig. \ref{fig:stelledineutroni} are, for each $N \leq  3$, comparable with the uncertainties in the SM. Results for the $N = 2, 3$ cases in the limit of equal masses for the $\chi_i$ are in agreement with both $\bar{\chi} n \rightarrow \chi\chi$ and neutron stars.
These analyses shed more light on the neutron anomaly and we conclude that the $n \rightarrow \chi\chi\chi$ decay model is predictive, satisfying
the experimental bounds on DM and neutron physics for $N > 1$.
\vspace{+1em}

\end{document}